\documentstyle[sprocl,epsf,array]{article}

\def\btod{$B \to D l \nu$}
\def\btods{$ B \to D^* l \nu$}
\def\btorho{ B^0\to\rho^+ l^- \nu_l}
\def\btopi{B^0\to\pi^+ l^- \nu_l}

\def\vub{|V_{ub}|}

\def\gev{\,\mathrm{Ge\kern-0.1em V}}
\def\mev{\,\mathrm{Me\kern-0.1em V}}

\newcommand{\err}[2]{%
{{\renewcommand{\arraystretch}{0.4}
\mathop{\raisebox{0.1\height}{\scriptsize
$\begin{array}{@{}c@{}}+\\-\end{array}$}}%
\raisebox{0.1\height}{\scriptsize
$\begin{array}{@{}r@{}}#1\\#2\end{array}$}}}%
}

\def\be{\begin{equation}}
\def\ee{\end{equation}}
\def\bea{\begin{eqnarray}}
\def\eea{\end{eqnarray}}

\begin{document}

\begin{titlepage}
\begin{flushright}
hep-ph/9709407
\end{flushright}
\vskip0.5cm
\begin{center}
{\Large\bf THE MOST PROMISING WAYS\\[5pt] TO MEASURE $V_{ub}$}

\vskip1cm

{\large Patricia Ball}

\vskip0.2cm

{\em Fermilab, P.O.\ Box 500, Batavia IL 60510, USA}

\vskip4cm

{\large\bf Abstract:\\[10pt]}
\parbox[t]{\textwidth}{
I review currently discussed methods to determine
the CKM mixing matrix element $|V_{ub}|$ from experimental
data. Although the theory of inclusive decays and their spectra has 
entered a model-independent stage, its predictions are still sensitive
to various input parameters, in particular the poorly known $b$ quark
mass. At present, determinations from exclusive channels, notably
$B\to\pi e \nu$, seem slightly more accurate and allow determination
of $|V_{ub}|$ from the spectrum in the momentum transfer to the
leptons with a theoretical uncertainty of $\sim 10\%$.}

\vfill

{\em To Appear in the Proceedings of the\\
7th International Symposium on Heavy Flavor Physics\\
Santa Barbara (CA), USA, 7--11 July 1997}
  
\end{center}
\end{titlepage}

\clearpage

\begin{titlepage}
\makebox[2cm]{\ }
\end{titlepage}

\clearpage

\setcounter{page}{1}

\title{THE MOST PROMISING WAYS TO MEASURE $V_{ub}$}

\author{ PATRICIA BALL }

\address{Fermilab, P.O.\ Box 500, Batavia IL 60510, USA}

\maketitle\abstracts{I review currently discussed methods to determine
the CKM mixing matrix element $|V_{ub}|$ from experimental
data. Although the theory of inclusive decays and their spectra has 
entered a model-independent stage, its predictions are still sensitive
to various input parameters, in particular the poorly known $b$ quark
mass. At present, determinations from exclusive channels, notably
$B\to\pi e \nu$, seem slightly more accurate and allow determination
of $|V_{ub}|$ from the spectrum in the momentum transfer to the
leptons with a theoretical uncertainty of $\sim 10\%$.}
  
\section{Introduction}

The study of $b\to u$ transitions will enter a new stage with the ever
increasing data available from CLEO and the new dedicated $B$
factories BABAR and BELLE and will eventually allow us to measure the
CKM mixing angle $|V_{ub}|$ accurately. Although most $b\to u$ are
purely hadronic, reliable theoretical predictions exist only for
semileptonic decays. Due to the progress of recent years in the
theoretical description of heavy quark decays, as
summarized e.g.\ in~\cite{mannel,vainshtein}, it is fair to say that
heavy quark physics has now reached the ``model-independent'' stage. I
would like to stress, however, that ``model-independent'' is not
equivalent to ``arbitrarily precise'': there is always
a sensitivity to input parameters like e.g.\ the $b$ quark mass, whose values
will always be afflicted with some theoretical and/or experimental 
uncertainty. ``Model-independence'' can rather be viewed as the fact that
attributing theoretical errors to predictions is now based on
more than simple guessing or averaging over several available models.

In the following I review the various suggested observables from which
$|V_{ub}|$ is likely to be measured reliably and where one can hope to
reach a minimum of the combined experimental and theoretical error.

\section{$|V_{ub}|$ from inclusive semileptonic decays}

The main tool in the theoretical description of inclusive $b$ decays
is heavy quark expansion.
As this topic is also discussed in Refs.~\cite{mannel,vainshtein}, I
will touch on it only shortly. It was realized in the
pioneering papers~\cite{bigi} that inclusive heavy
hadron decays can be described in terms of an expansion of the relevant
hadronic matrix element in inverse powers of the heavy quark mass. The
first term in this expansion is the free quark decay contribution and
the first correction term is suppressed by two powers of the $b$ quark
mass (at least in the total rate). Thus the first nonperturbative
corrections are small, of the order of 5\%, and the decay rate of the
$B$ meson nearly equals that of the $b$ quark. 
Heavy quark expansion works best in regions of phase-space
where the final-state hadrons carry large energy and it breaks
down if only a few resonances contribute.

Despite the smallness of nonperturbative power-suppressed corrections,
nonperturbative quantities also enter from a different source, namely
in form of the $b$ quark mass, which is related to the $B$ meson mass
by heavy quark expansion. Despite much effort to determine the $b$
quark mass, e.g.~\cite{bmass}, it is fair to say that this quantity
constitutes today one of the main sources of theoretical uncertainty
in inclusive decays.

Due to the overwhelming dominance of $b\to c$ transitions, the total
rate $\Gamma(B\to X_u e \nu)$ is not accessible in experiment. One
thus aims to measure spectra in one (or more) variables. Two
``natural'' variables in that game are the charged lepton energy 
 and the hadronic mass in the final
state. Experimental data~\cite{cleoinclusive} exist so far only for 
$d\Gamma/dE_e$ with lepton energies $E_e$ above about 2.3$\,$GeV. The cut
removes up to about 90\% of all events. The theoretical description of
the spectrum is very difficult in this region where fixed order heavy
quark expansion breaks down. A solution is 
to resum terms of all orders in $1/m_b$ into a so-called shape
function~\cite{minngang,neubert}, which is not known from first
principles, but in some rather remote future may be measurable from the
photon-energy spectrum in $B\to X_s\gamma$.

\begin{figure}
\makebox[1cm]{\ }\\[-3.8cm]
\epsfysize=.7\textheight
\centerline{\epsffile{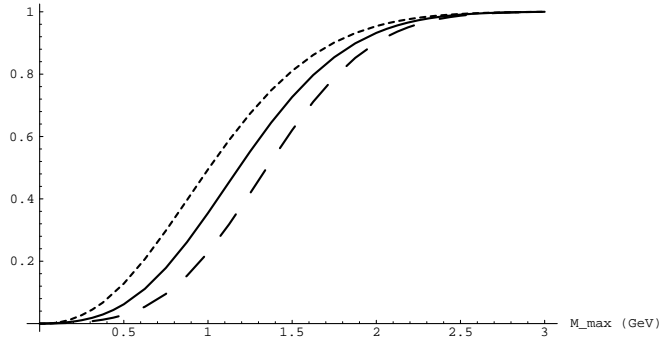}}
\vspace{-4cm}
\caption[]{
The integrated fraction of the events $\Phi(m_x^{\rm cut})$, 
Eq.~(\protect{\ref{eq:x}}), for different values of the $b$ quark mass,
$m_b=\{4.72,4.82,4.92\}\,$GeV. Figure taken from Ref.~\protect{\cite{keys}}.
}\label{fig:1}
\end{figure}

Recently, two groups took up an older suggestion~\cite{barger} and
studied the spectrum in the invariant hadronic mass $m_X$ 
\cite{FLW,keys}. The region of small hadronic multiplicity corresponds
here to small $m_X$, the threshold for $b\to c$ transitions is at
$m_X=m_D$. A cut at $m_X=m_D$ removes only a moderate fraction 
of $b\to\ u$ transitions. There is an experimental problem to be
expected, though, the ``charm-leaking'' of misidentified charmed
particles below the kinematical $b\to c$ threshold. In order to remove
them effectively, it may be necessary to cut off the hadron mass
spectrum at smaller values of $m_X$, say $m_X\approx 1.5\,$GeV. As, on
the other hand, in a fixed order heavy quark expansion one also has to
cut off the contributions of small $m_X$, say $m_X\leq 1\,$GeV, the
resulting bin in hadronic mass may be too small for heavy quark
expansion to be applicable~\cite{FLW}. Like for the endpoint spectrum in
the electron energy, it may be necessary to invoke the shape-function,
which is known only via its first few moments. Definitive predictions
thus almost necessarily involve a certain degree of model-dependence
\cite{FLW,keys}. 

In~\cite{keys} the integrated fraction of events was
introduced,
\begin{equation}\label{eq:x}
\Phi(m_x^{\rm cut}) = \frac{1}{\Gamma(B\to X_u e
  \nu)}\,\int_0^{m_X^{\rm cut}} dm_X\,\frac{d\Gamma}{dm_X},
\end{equation}
and studied in its sensitivity to its input parameters, in particular
the $b$ quark mass. In Fig.~\ref{fig:1} I show $\Phi(m_X^{\rm cut})$
as function of $m_X$ for various values of $m_b$. It is clear that the
strong sensitivity to $m_b$ around $m_X^{\rm cut}=1.5\,$GeV is not 
favourable to extracting
$|V_{ub}|$. There are also other theoretical uncertainties
not shown in the plot. The authors of~\cite{keys} quote a theoretical
uncertainty of $|V_{ub}|$ of about (10--20)\% for the pessimistic
scenario $m_X^{\rm cut}=1.5\,$GeV. Both papers~\cite{FLW,keys} agree
that the main source of uncertainty is the value of $m_b$ or
$\bar\Lambda = m_B-m_b+O(1/m_b)$, respectively, which needs to be fixed to an
accuracy of 100$\,$MeV or better in order to reduce the theoretical
error on $|V_{ub}|$ arising from $m_b$ to 10\%. The total theoretical
error should also account for the model-dependence introduced by the
specific choice of the shape-function and for subleading
``higher-twist'' effects and may be closer to the 20\% mark. Heavy
quark expansion to fixed order without introducing the shape-function
is only sensible if $m_X^{\rm cut}$ can be pushed to larger
values. But also in this case there is a strong dependence of the
result on $\bar\Lambda$ \cite{FLW}.

\section{$|V_{ub}|$ from exclusive semileptonic decays}

Possible candidates for the extraction of $|V_{ub}|$ from exclusive
decays are $B\to\pi e \nu$, $B\to \rho e \nu$ and $B\to\omega
e\nu$. CLEO has already measured the corresponding
rates~\cite{gibbons}, but the results are still slightly
model-dependent. Theory has to provide the form factors that describe
the relevant hadronic matrix elements, which can be parameterized in
the following form:
\begin{eqnarray}
\langle\pi|V_\mu|B\rangle & = &
f_+(q^2)(p_B+p_\pi)_\mu+\dots\nonumber\\
\langle\rho|(V-A)_\mu|B\rangle & = &
-i(m_B+m_\rho)A_1(q^2)\epsilon^*_\mu +
\frac{iA_2(q^2)}{m_B+m_\rho}\,(\epsilon^*\cdot
p_B)(p_B+p_\rho)_\mu\nonumber\\
& & {} +
\frac{2V(q^2)}{m_B+m_\rho}\,\epsilon_{\mu\nu\alpha\beta}
\epsilon^{*\nu} p_B^\alpha p_\rho^\beta+\dots
\end{eqnarray} 
The form factors denoted by dots do not contribute to semileptonic
 decays with massless leptons. The form factors are functions of
 $q^2$, the squared momentum transfer to the leptons. 
Reliable predictions for form factors come from both lattice
calculations and light-cone sum rules; there exist also a number of
useful parametrizations and constraints.

\renewcommand{\arraystretch}{1.2}
\begin{table}[t]
\begin{center}
\hbox to\hsize{\hss
\begin{tabular}{l>{$}l<{$}>{$}l<{$}>{$}l<{$}>{$}l<{$}>{$}l<{$}>{$}l<{$}}
\hline
 & f_+(0) & A_1(0) & A_2(0) & V(0)\\
\hline
UKQCD~\cite{para} & 0.27\pm0.11 & 0.27\err{0.05}{0.04} & 0.26\err{0.05}{0.03}
  & 0.35\err{0.06}{0.05} &\\
LCSR~\cite{brho} &
 & 0.27\pm0.05 & 0.28\pm0.05 &  0.35\pm0.07 \\
LCSR~\cite{bpi} & 0.28\pm 0.05\\
\hline
\end{tabular}\hss}
\end{center}
\bigskip
\addtolength{\arraycolsep}{-1pt}
\begin{center}
\hbox to\hsize{\hss
\begin{tabular}{l>{$}l<{$}>{$}l<{$}>{$}l<{$}>{$}l<{$}}
\hline
 & \Gamma(\btopi) & \Gamma(\btorho) & \Gamma(\rho)/\Gamma(\pi) &
 \Gamma_L/\Gamma_T\\
\hline
UKQCD~\cite{para}  & 8.5\err{3.3}{1.4} & 16.5\err{3.5}{2.3} & 
1.9\err{0.9}{0.7} &
0.80\err{0.04}{0.03} \\
LCSR~\cite{brho} &
  & 13.5\pm4.0 & 1.7\pm0.5 & 0.52\pm0.08\\
LCSR~\cite{bpiold} &
  8.7\pm 2.6 \\
\hline
\end{tabular}\hss}
\end{center}
\addtolength{\arraycolsep}{1pt}
\caption[]{Form factor values at $q^2 =0$ and decay rates and ratios
for $b\to u$ transitions from lattice-constrained parametrizations and
from light cone sum
rules (LCSR). Decay rates are given in units of $\vub^2
\,\mathrm{ps}^{-1}$.}
\label{tab:compare}
\renewcommand{\arraystretch}{1}
\end{table}

Let me first shortly review lattice results, which are
discussed in detail in~\cite{flynn}. 
At present, direct calculation of form factors
from lattice~\cite{UKQCD} is possible only for large $q^2\geq
14\,$GeV$^2$. Recently, however, the UKQCD collaboration has designed
a simple parametrization for form factors that describe the decay of a
$B$ meson into a light meson~\cite{para}. This parametrization is
inspired by the work of Stech~\cite{stech} and consistent with heavy
quark symmetry and kinematical constraints, but requires an ansatz for
the $q^2$ dependence of one of the form factors. The parameters of
this ansatz are determined by fitting to lattice results around
$q^2_{\rm max}$. As a result, $B\to\rho e\nu$ and $B\to K^*\gamma$
decays are described with only two parameters and $B\to\pi e\nu$
decays with a further two. The form factors and decays rates are given
in Table~1. The resulting spectra are shown in~\cite{flynn,para}. 
The uncertainties for
$f_+$ and the $B\to\pi$ spectra are still rather large, whereas the
uncertainties in the spectrum of $B\to\rho e\nu$ are less than 20\%
for large $q^2\geq 15\,$GeV$^2$ and thus could allow a measurement of
$|V_{ub}|$ with a theoretical error of about 10\%.


Progress in describing the shape of form factors
has recently been made in the form of model-independent 
parameterizations~\cite{bgl2,savage} based on QCD dispersion
relations and analyticity. These dispersion relations  
lead to an infinite tower of upper and lower
bounds that can be derived by using 
the normalizations of the form factor $f_+(q^2_i)$
at a fixed number of kinematic points $q^2_i$ as 
input~\cite{brm,bgl1}. 
In Ref.~\cite{bgl2,savage} the most general parametrization of a form
factor consistent with the constraints from QCD was derived.
For a generic form factor $F(q^2)$ describing
the exclusive semileptonic decay of a $B$ meson to a final
state meson $H$ as a function of $q^2$, 
the parameterization takes the form 
\begin{eqnarray}\label{master}
F(q^2) = {1\over P(q^2) \phi(q^2)} \sum_{k=0}^\infty a_k\  z(q^2;q^2_0)^k 
\ \ \ \ ,
\end{eqnarray}
where $\phi(q^2)$ is a computable function arising from 
perturbative QCD.  The function $P(q^2)$ depends only on
the masses of mesons below the $BH$ 
pair-production threshold
that contribute to $BH$ pair-production as virtual
intermediate states.
The variable $z(q^2;q^2_0)$ is a kinematic function of $q^2$ defined by
\begin{eqnarray}\label{zdef}
{1+z(q^2;q^2_0) \over 1-z(q^2;q^2_0)} = \sqrt{ q^2_+ -q^2 \over q^2_+ - q^2_0}
\ \ \ \ \ ,
\end{eqnarray}
where $q^2_+ = (M_B + M_H)^2$ is the pair-production threshold
and $q^2_0$ is a free parameter that is 
often  taken to be
$q^2_- = (M_B - M_H)^2$, the maximum momentum-transfer squared
allowed in the semileptonic decay $B \to H l \nu$.
The coefficients $a_k$ are unknown constants constrained to obey
\begin{eqnarray}\label{asum}
\sum_{k=0}^\infty \left( a_k \right)^2 \leq 1
\ \ \ \  .
\end{eqnarray}
The kinematic function $z(q^2;q^2_0)$ takes its minimal 
physical value $z_{min}$ 
at $q^2=q^2_-$, vanishes at $q^2=q^2_0$, and reaches its
maximum $z_{max}$ at $q^2=0$.
Thus the sum $\sum a_k\ z^k$ is a series expansion about 
the kinematic point $q^2=q^2_0$. The value $z_{max}$
can be made even smaller by choosing an
optimized value $0 \le q^2_0 \le q^2_-$.
In that case, most form factors describing \btod\
and \btods\ can be parameterized with only one unknown constant
to an accuracy of a few percent (assuming the normalization at zero 
recoil given by heavy quark symmetry).
Thus the continuous function $F(q^2)$ has been reduced to a single
constant, for example the value of the form factor $F(q^2=0)$
at maximum recoil. For $\bar B \to \pi l \overline \nu$, the maximum
value of $z$ is $z_{max} = 0.52$, but even in this case 
Eqs.~(\ref{master})\
and (\ref{asum})\
severely constrain the relevant form factor~\cite{bgl1}. 
The main source of uncertainty in this approach is the
normalization of the form factor which has to be taken from an external
source, such that the uncertainty of the predicted form factor is at
least as large as the error on the input normalization. In
Ref.~\cite{savage} 30\% are quoted.

\begin{figure}
\epsfxsize=200pt
\centerline{\epsffile{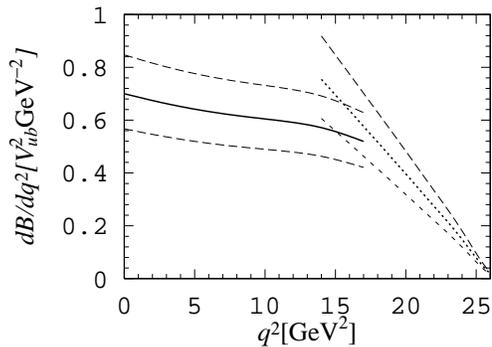}}
\caption[]{The spectrum $dB(B\to\pi e\nu)/dq^2$  as a function of
  $q^2$. Solid line: result from light-cone sum rules. Dotted line:
  $B^*$ contribution in pole dominance approximation, which is
  expected to dominate for large $q^2$. Dashed lines: 
theoretical uncertainties. Figure
  taken from \protect{\cite{bpi}}.}\label{fig:bpi}
\end{figure}

The last method I would like to discuss in these proceedings are the
so-called light-cone sum rules~\cite{LC}, which were applied to $B\to\pi$ and
$B\to\rho$ transitions in~\cite{bpiold,bpi,brho,PB93}. The starting point
in this approach is the observation that at
large recoil the light quark originating
from the weak decay carries large energy of order $m_b/2$ and
has to transfer it to the soft cloud to recombine to the final 
state hadron. 
The probability of such a recombination depends on the parton content
of both the $B$ meson and the light meson, the valence configuration with
the minimum number of Fock constituents 
being dominant. 
The valence quark configuration is characterized by the wave function
$\phi(x,k_\perp)$ depending on the momentum fraction $x$ carried by the
quark and on its transverse momentum $k_\perp$. There exist two different 
mechanisms for the valence quark contribution to the transition form 
factor.
The first one is the hard rescattering mechanism which requires that
the recoiling and spectator quarks are at small transverse separations.
In this case the large momentum is transferred by exchange
of a hard gluon with virtuality $k^2\sim O(m_b)$. This contribution is 
perturbatively calculable in terms of the 
Bethe-Salpeter wave functions at 
small ($\sim 1/m_b$) transverse separations, or 
{\em distribution amplitudes} (DA):
\begin{equation}
\phi(x) =\int^{k_\perp^2\sim m_b}dk_\perp^2
\phi(x,k_\perp).
\end{equation}
The second mechanism is the soft contribution.
The idea is that hard gluon exchange is not necessary, provided 
one picks up an ``end-point'' configuration with almost all 
momentum $1-x\sim O(1/m_b)$ carried by one constituent.
The transverse quark-antiquark separation is not constrained
in this case, which implies that the soft contribution is sensitive 
to long-distance dynamics. To calculate the soft
contribution one needs to know the wave function as a function of the
transverse separation; the simpler distribution amplitude is not enough.
QCD sum rules offer a nonperturbative
technique to estimate the necessary convolution integral without
explicit knowledge of the wave functions.
\begin{table}
\renewcommand{\arraystretch}{1.2}
\addtolength{\arraycolsep}{3pt}
$$
\begin{array}{l|llr}
{\rm FF} & F(0) & \phantom{-}a_F & \multicolumn{1}{l}{\phantom{-}b_F}\\ \hline
f_+^{B\to\pi} & 0.30\pm 0.03 & -1.32 & 0.21\\
f_0^{B\to\pi} & 0.30 \pm 0.05 & -0.84 & 0.03\\
A_1^{B\to\rho} & 0.27\pm 0.05 & -0.42 & -0.29\\
A_2^{B\to\rho} & 0.28\pm 0.05 & -1.34 & 0.38\\
V^{B\to\rho} & 0.35\pm 0.07 & -1.51 & 0.47
\end{array}
$$
\caption[]{Form factors from light-cone sum rules with
functional $q^2$ dependence
fitted to Eq.~(\ref{eq:tdependence}).}\label{table:formfactorsLC}
\renewcommand{\arraystretch}{1}
\addtolength{\arraycolsep}{-3pt}
\end{table}

\begin{figure}
\epsfxsize=\textwidth
\centerline{\epsffile{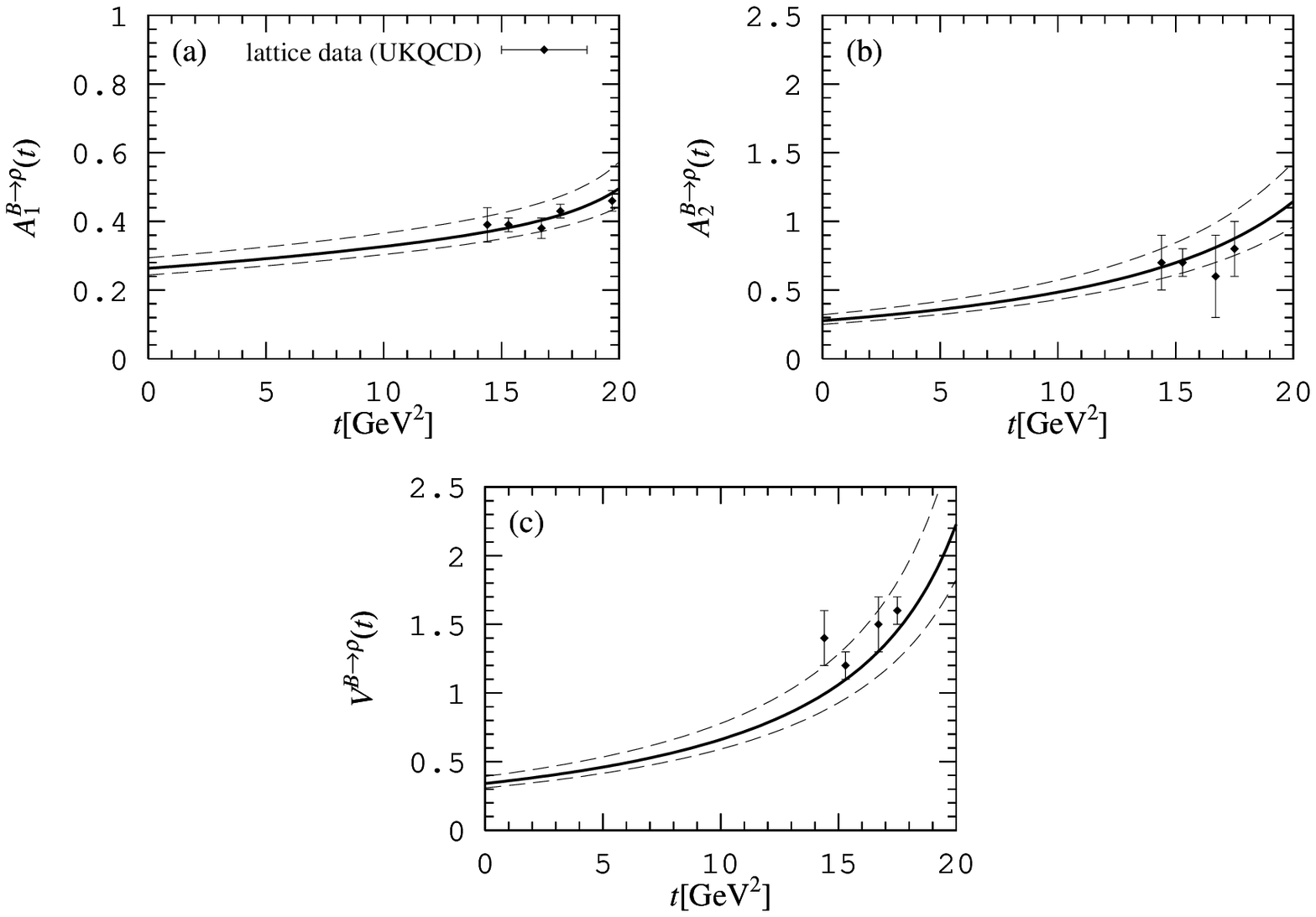}}
\caption[]{Form factors of the $B\to\rho$ transition from light-cone
  sum rules \protect{\cite{brho}} (solid lines). Dashed lines:
  estimate of theoretical errors. For comparison I also plot the
  results from lattice simulations \protect{\cite{UKQCD}}. Figure
  taken from \protect{\cite{brho}}.}\label{fig:Brho}
\vspace*{1cm}
\epsfxsize=\textwidth
\centerline{\epsffile{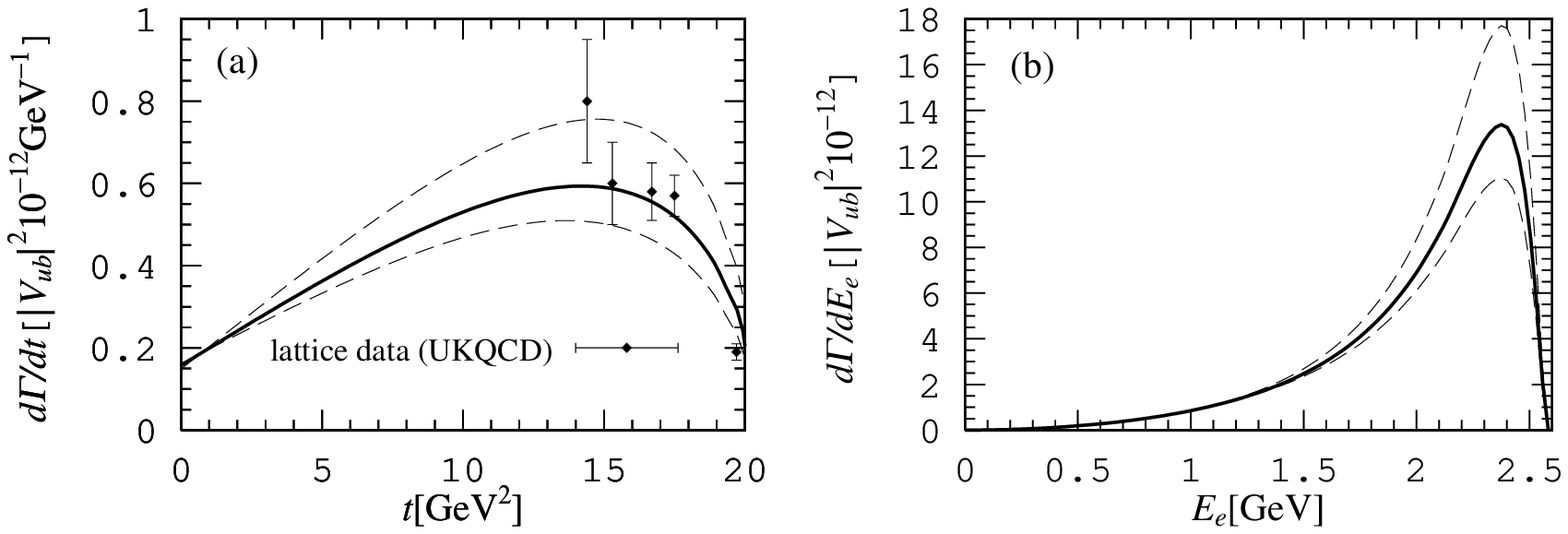}}
\caption[]{Spectra of the $B\to\rho$ decay in the momentum transfer
  (a) and the electron energy (b). Same notations as in previous
  figure.}\label{fig:Brhospec}
\end{figure}

The essential nonperturbative input in this method is encoded in
hadronic distribution amplitudes ordered by increasing twist. The
lowest order twist~2 distributions can be experimentally accessed,
e.g.\ in the $\pi\gamma\gamma^*$ form factor at large momentum
transfer. Lacking this measurement, the most important nonperturbative
terms in the DAs have been estimated from QCD sum rules~\cite{rhoWF}.
Light-cone sum rules are expected to be valid for not too large $q^2$,
$m_b^2-q^2\sim O(m_b)$, e.g.\ $q^2\leq 17\,$GeV$^2$, and in that
respect are largely complementary
to presently feasible lattice simulations. At present, the
calculations for $f_+(q^2)$ include twist 3 and 4 distributions and 
lowest order radiative corrections to the leading twist contribution 
\cite{bpi}. Both the twist expansion and the radiative corrections are
well under control. The spectrum $d\Gamma(B\to\pi e\nu)/dq^2$ is shown in
Fig.~\ref{fig:bpi}. The method of light-cone sum rules cannot yield
arbitrarily accurate results, but always involves a certain systematic
error, which cannot be reduced to below $\sim 10\,$\% in the form
factor. In this regard the results of~\cite{bpi} are not expected to
be much improved by future calculations.

The situation is different for the $B\to\rho$ transition, where so
far only the tree-level twist 2 contributions have been taken into
account~\cite{brho}. Results for the form factor are shown in
Fig.~\ref{fig:Brho}, for the spectra in Fig.~\ref{fig:Brhospec}. 
Fig.~\ref{fig:Brho} strikes by the excellent agreement between LCSR
and lattice results, cf.\ Table 1. However, it may be premature to conclude
that the question of $B\to\rho$ form factors is settled, as the LCSR
results still need to be improved by including radiative
and higher twist corrections. Results for form factors and branching
ratios for both $B\to\pi$ and $B\to\rho$ transitions are given in
Table~1. In Table~2 I also give a simple parametrization of LCSR form
factors of the form
\begin{equation}
\label{eq:tdependence}
F(q^2) = \frac{F(0)}{1+a_F\, \frac{q^2}{m_B^2} + b_F\, \frac{q^4}{m_B^4}}.
\end{equation}

\section{Conclusions}

In my opinion $|V_{ub}|$ will finally be determined from the spectrum
$d\Gamma(B\to\pi e \nu)/dq^2$. As I have discussed in Sec.~2, the
connection between theory and experiment in inclusive decays may be
difficult to establish, in particular if the experimental cut-off in
the hadronic invariant mass has to be pessimistically low. In any
case one has to await runing and data-taking at BABAR or
BELLE. Exclusive decays, on the other hand, are already measured at
CLEO with increasing statistics. {}From the experimental point of view,
the $B\to\rho e\nu$ decay is slightly disfavoured, as distinguishing 
nonresonant $\pi\pi$ states
from the broad $\rho$ resonance poses an additional experimental
challenge that is absent in $B\to\pi e\nu$. On the other hand, the
branching 
ratio of the
latter one is smaller by roughly a factor of two. Theory provides a
number of largely different and complementary theoretical tools which
become continually finer shaped. Also for theory $B\to\rho$ is more
challenging, as three form factors need to be predicted, and to date
the predicted $q^2$ dependence can not yet be checked for internal consistency 
 from unitarity constraints as it is the case with
$B\to\pi$.\footnote{The parametrizations suggested in~\cite{para} are
  not as strict as the ones from unitarity.} In principle it is of
course possible to constrain the $q^2$ dependence experimentally by
measuring the polarization of the $\rho$, but in view of the lack of
conclusive results for the Cabibbo-favoured decays $D\to K^* e\nu$ and
even $B\to\ D^* e \nu$, this possibility appears remote. As none of the
discussed methods can predict the form factor in the complete physical
range in $q^2$, a determination of $|V_{ub}|$ from the broad spectrum
of $B\to\pi e\nu$ in $q^2$ seems most promising and also allows
naturally the inclusion of experimental cuts.


\section*{References}
\begin{thebibliography}{99}

\bibitem{mannel} T. Mannel, {\em these proceedings}.

\bibitem{vainshtein} A. Vainshtein, {\em these proceedings}.

\bibitem{bigi} I. Bigi et al., Phys.\ Lett.\ B {\bf 293} (1992) 430;
  Err.\ ibid.\ B {\bf 297} (1993) 477; Phys.\ Rev.\ Lett.\ {\bf 71} (1993) 496.

\bibitem{keys} R.D.\ Dikeman and N.G.\ Uraltsev, hep--ph/9703437;\\
I. Bigi, R.D.\ Dikeman and N.G.\ Uraltsev, hep--ph/9706520.

\bibitem{bmass} P. Ball and V.M.\ Braun, Phys.\ Rev.\ D {\bf 49}
  (1994) 2472;\\ 
M. Jamin and A. Pich, hep--ph/9702276.

\bibitem{cleoinclusive} CLEO coll., Phys.\ Rev.\ Lett.\ {\bf 71} (1993) 4111.

\bibitem{minngang} I. Bigi et al., Int.\ J. Mod.\ Phys.\ A {\bf 9}
  (1994) 2467;\\
R.D.\ Dikeman, M. Shifman and N.G.\ Uraltsev, Int.\ J. Mod.\ Phys.\ A
  {\bf 11} (1996) 571.

\bibitem{neubert} M. Neubert, Phys.\ Rev.\ D {\bf 49} (1994) 3392; 
Phys.\ Rev.\ D {\bf 49} (1994) 4623.

\bibitem{barger} V. Barger, C.S.\ Kim and G.J.N.\ Phillips, Phys.\
  Lett.\ B {\bf 251} (1990) 629.

\bibitem{FLW} A.F.\ Falk, Z. Ligeti and M.B.\ Wise, Phys.\ Lett.\ B
  {\bf 406} (1997) 225.

\bibitem{gibbons} L. Gibbons, {\em these proceedings}

\bibitem{para} UKQCD coll., hep--lat/9708008.

\bibitem{brho} P. Ball and V.M.\ Braun, Phys.\ Rev.\ D {\bf 55} (1997)
  5561.

\bibitem{bpi}A. Khodjamirian {\em et al.}, hep-ph/9706303;\\
E. Bagan {\em et al.}, hep-ph/9709243.

\bibitem{bpiold} V.M.\ Belyaev, A. Khodjamirian and  R. R\"{u}ckl,
Z. Phys.\ C {\bf 60} (1993) 349;
A. Khodjamirian and  R. R\"{u}ckl, Talk at ICHEP 96, Warsaw,
Poland, July 1996; published in  ICHEP 96, 902 (hep--ph/9610367).

\bibitem{flynn} J. Flynn, {\em these proceedings}.

\bibitem{UKQCD} 
UKQCD coll., Nucl.\ Phys.\ {\bf B461} (1996) 327;
Nucl.\ Phys.\ {\bf B476} (1996) 313.

\bibitem{stech} B. Stech, Phys.\ Lett.\ B {\bf 354} (1995) 447.

\bibitem{bgl2} G. Boyd, B. Grinstein and R. Lebed, Phys.\ Lett.\ B
  {\bf 353} (1995) 306; Nucl.\ Phys.\ {\bf B461} (1996) 493.

\bibitem{savage}
C. Boyd and M. Savage, Phys.\ Rev.\ D {\bf 56} (1997) 303.

\bibitem{brm} C. Bourrely, B. Machet and E. de Rafael, Nucl.\ Phys.\
  {\bf B189} (1981) 157.

\bibitem{bgl1} C. Boyd, B. Grinstein and R. Lebed, Phys.\
  Rev.\ Lett.\ {\bf 74} (1995) 4603;\\
L. Lellouch, Nucl.\ Phys.\ {\bf B479} (1996) 353.

\bibitem{LC} P. Ball, V.M.\ Braun and H.G.\ Dosch, Phys.\ Rev.\ D {\bf
 44} (1991) 3567;\\
 A. Ali, V.M.\ Braun and H. Simma, Z.\ Phys.\ C {\bf 63}
  (1994) 437.

\bibitem{PB93} P. Ball, Phys.\ Rev.\ D {\bf 48} (1993) 3190 and
  references therein.

\bibitem{rhoWF}  V.M.\ Braun and I.E.\ Filyanov, Z.\ Phys.\ C {\bf {44}}
  (1989) 157;\\
P. Ball and V.M.\ Braun, Phys.\ Rev.\ D {\bf 54} (1996) 2182.

\end{thebibliography}
\end{document}